\documentclass[twocolumn,]{aastex63}

\newcommand{\target}{MACS0416\_Y1}
\newcommand{\oiii}{[\ion{O}{3}]}
\newcommand{\cii}{[\ion{C}{2}]}
\newcommand{\hii}{\ion{H}{2}}
\newcommand{\kms}{km~s$^{-1}$}
\newcommand{\hst}{\textit{HST}}
\newcommand{\hubble}{\textit{Hubble}}
\newcommand{\jwst}{\textit{JWST}}

\received{}
\revised{}
\accepted{}

\shorttitle{300 pc imaging of a $z=8.3$ galaxy}
\shortauthors{Tamura et al.}

\begin{document}

\title{The 300 pc resolution imaging of a $z$ = 8.31 galaxy:
Turbulent ionized gas and potential stellar feedback 600 million years after the Big Bang}


\correspondingauthor{Yoichi Tamura}
\email{ytamura@nagoya-u.jp}

\author[0000-0003-4807-8117]{Yoichi Tamura}
\affiliation{Department of Physics, Graduate School of Science, Nagoya University, Furo, Chikusa, Nagoya, Aichi 464-8602, Japan}
\author[0000-0002-5268-2221]{Tom J.\ L.\ C.\ Bakx}
\affiliation{Department of Physics, Graduate School of Science, Nagoya University, Furo, Chikusa, Nagoya, Aichi 464-8602, Japan}
\affiliation{National Astronomical Observatory of Japan, 2-21-1, Osawa, Mitaka, Tokyo 181-8588, Japan}
\author[0000-0002-7779-8677]{Akio K.\ Inoue}
\affiliation{Department of Pure and Applied Physics, Graduate School of Advanced Science and Engineering, Faculty of Science and Engineering, Waseda University, 3-4-1, Okubo, Shinjuku, Tokyo 169-8555, Japan}
\affiliation{Waseda Research Institute for Science and Engineering, Faculty of Science and Engineering, Waseda University, \\3-4-1, Okubo, Shinjuku, Tokyo 169-8555, Japan}
\author[0000-0002-0898-4038]{Takuya Hashimoto}
\affiliation{Tomonaga Center for the History of the Universe (TCHoU), Faculty of Pure and Applied Sciences, University of Tsukuba, \\Tsukuba, Ibaraki 305-8571, Japan}
\author{Tsuyoshi Tokuoka}
\affiliation{Department of Pure and Applied Physics, Graduate School of Advanced Science and Engineering, Faculty of Science and Engineering, Waseda University, 3-4-1, Okubo, Shinjuku, Tokyo 169-8555, Japan}
\author{Chihiro Imamura}
\affiliation{Department of Physics, Graduate School of Science, Nagoya University, Furo, Chikusa, Nagoya, Aichi 464-8602, Japan}
\author[0000-0001-6469-8725]{Bunyo Hatsukade}
\affiliation{National Astronomical Observatory of Japan, 2-21-1, Osawa, Mitaka, Tokyo 181-8588, Japan}
\affiliation{Institute of Astronomy, Graduate School of Science, The University of Tokyo, 2-21-1, Osawa, Mitaka, Tokyo 181-0015, Japan}
\author[0000-0002-2419-3068]{Minju M.\ Lee}
\affiliation{Cosmic Dawn Center (DAWN), Jagtvej 128, DK-2200 Copenhagen N, Denmark}
\affiliation{DTU-Space, Technical University of Denmark, Elektrovej 327, DK-2800 Kgs.\ Lyngby, Denmark}
\affiliation{Max-Planck-Institut f\"{u}r Extraterrestrische Physik (MPE), Giessenbachstr. 1, D-85748 Garching, Germany}
\author[0000-0003-3349-4070]{Kana Moriwaki}
\affiliation{Research Center for the Early Universe, Graduate School of Science, The University of Tokyo, 7-3-1 Hongo, Bunkyo, Tokyo 113-0033, Japan}
\affiliation{Department of Physics, Graduate School of Science, The University of Tokyo, 7-3-1 Hongo, Bunkyo, Tokyo 113-0033, Japan}
\author[0000-0003-0137-2490]{Takashi Okamoto}
\affiliation{Department of Cosmosciences, Graduate School of Science, Hokkaido University, N10W8, Kitaku, Sapporo 060-0810, Japan}
\author[0000-0002-7675-5923]{Kazuaki Ota}
\affiliation{Frontier Science and Social Co-creation Initiative, Kanazawa University, Kakuma-machi, Kanazawa, Ishikawa 920-1192 Japan}
\author[0000-0003-1937-0573]{Hideki Umehata}
\affiliation{Institute for Advanced Research, Nagoya University, Furo, Chikusa, Nagoya, Aichi 464-8601, Japan}
\affiliation{Department of Physics, Graduate School of Science, Nagoya University, Furo, Chikusa, Nagoya, Aichi 464-8602, Japan}
\author[0000-0001-7925-238X]{Naoki Yoshida}
\affiliation{Department of Physics, Graduate School of Science, The University of Tokyo, 7-3-1 Hongo, Bunkyo, Tokyo 113-0033, Japan}
\author[0000-0003-1096-2636]{Erik Zackrisson}
\affiliation{Observational Astrophysics, Department of Physics and Astronomy, Uppsala University, Box 516, SE-751 20 Uppsala, Sweden}
\author[0000-0001-8083-5814]{Masato Hagimoto}
\affiliation{Department of Physics, Graduate School of Science, Nagoya University, Furo, Chikusa, Nagoya, Aichi 464-8602, Japan}
\author[0000-0003-3278-2484]{Hiroshi Matsuo}
\affiliation{National Astronomical Observatory of Japan, 2-21-1, Osawa, Mitaka, Tokyo 181-8588, Japan}
\author{Ikkoh Shimizu}
\affiliation{Department of Literature, Shikoku Gakuin University, 3-2-1 Bunkyocho, Zentsuji, Kagawa 765-8505, Japan}
\author[0000-0001-6958-7856]{Yuma Sugahara}
\affiliation{Waseda Research Institute for Science and Engineering, Faculty of Science and Engineering, Waseda University, \\3-4-1, Okubo, Shinjuku, Tokyo 169-8555, Japan}
\affiliation{National Astronomical Observatory of Japan, 2-21-1, Osawa, Mitaka, Tokyo 181-8588, Japan}
\author[0000-0001-8416-7673]{Tsutomu T.\ Takeuchi}
\affiliation{Department of Physics, Graduate School of Science, Nagoya University, Furo, Chikusa, Nagoya, Aichi 464-8602, Japan}
\affiliation{The Research Center for Statistical Machine Learning, the Institute of Statistical Mathematics, 10-3 Midori-cho, Tachikawa, Tokyo 190---8562, Japan}

\begin{abstract}

We present the results of 300~pc resolution ALMA imaging of the \oiii{} 88~$\micron$ line and dust continuum emission from a $z = 8.312$ Lyman break galaxy MACS0416\_Y1.
The velocity-integrated \oiii{} emission has three peaks which are likely associated with three young stellar clumps of MACS0416\_Y1, while the channel map shows a complicated velocity structure with little indication of a global velocity gradient unlike what was found in \cii{} 158~$\micron$ at a larger scale, suggesting random bulk motion of ionized gas clouds inside the galaxy. In contrast, dust emission appears as two individual clumps apparently separating or bridging the \oiii{}/stellar clumps. The cross correlation coefficient between dust and ultraviolet-related emission (i.e., \oiii{} and ultraviolet continuum) is unity on a galactic scale, while it drops at $< 1$~kpc, suggesting well mixed geometry of multi-phase interstellar media on sub-kpc scales.
If the cutoff scale characterizes different stages of star formation, the cutoff scale can be explained by gravitational instability of turbulent gas.
We also report on a kpc-scale off-center cavity embedded in the dust continuum image. This could be a superbubble producing galactic-scale outflows, since the energy injection from the 4~Myr starburst suggested by a spectral energy distribution analysis is large enough to push the surrounding media creating a kpc-scale cavity.

\end{abstract}

\keywords{Galaxy evolution (594) --- Galaxy formation (595) --- Interstellar medium (847) --- Stellar feedback (1602) --- Superbubbles (1656)}


\section{Introduction} \label{sec:introduction}

What young galaxies in the epoch of reionization (EoR) look like is one of the most fundamental questions in astronomy. A recent study based on the samples of $z \gtrsim 8$ Lyman-break galaxies (LBGs) has reported a rapid rise of the ultraviolet (UV) luminosity function by one order of magnitude from $z \sim 10$ to $\sim 8$ \citep{Oesch18, Bouwens21, Bouwens23a, Bouwens23b}, implying strong evolution of the cosmic star-formation rate density within a short duration of time ($\lesssim 200$~Myr) while no strong evolution at the bright end is also suggested (e.g., \citealt{Morishita18, Finkelstein22, Naidu22, Donnan23, Harikane22b}, see also \citealt{Harikane22}). For the former case the steep evolution can be attributed to mergers and accretion of baryonic matter, whereas the latter requires extremely high star-formation efficiencies to form such massive systems in a limited cosmic age. In either case, the physical process and properties of stars and interstellar medium (ISM) are closely related to what drives the rapid evolution of early galaxies.

The dynamically-unrelaxed nature of stars and ISM is, however, bound to create complicated morphological and physical properties of young galaxies. Furthermore, newly-born stars and subsequent type-II supernova (SN II) explosions affect the surrounding media radiatively, mechanically, and chemically. This does not only alter the morphology of these galaxies, but changes the chemical and ionization states of the ISM \citep[e.g.,][]{Hirashita02}. For most galaxies found in the EoR, only global properties, such as star-formation rate (SFR), stellar and gas masses, have been constrained thus far, although recent ALMA observations have revealed different distributions of neutral and ionized ISM in $z \sim 7$ UV-selected galaxies \citep[e.g.,][]{Inoue16, Carniani17, Witstok22, Akins22, Algera23}. Investigating spatially-resolved properties of ISM in the heart of the EoR ($z \sim 8$) is an important next step to understand the fast evolution of galaxies.

Recently, a bright ($H_{160} \approx 25.9$) $Y$-band dropout LBG, \target{} \citep[e.g.,][]{Infante15, Laporte15}, has been identified with ALMA both in the far-infrared (FIR) \oiii{} 88~$\micron$ and \cii{} 158~$\micron$ lines, yielding a spectroscopic redshift of $z=8.312$ \citep{Tamura19, Bakx20}. This LBG lies behind a \hubble{} Frontier Field (HFF) cluster, MACS J0416.1$-$2403, while the gravitational lensing is moderate with a magnification factor of $\mu_{\rm g} = 1.43 \pm 0.04$ \citep{Kawamata16}. The \oiii{}-to-\cii{} luminosity ratio is high ($\approx 9.3 \pm 2.6$, \citealt{Bakx20}, see also \citealt{Carniani20}) compared with local star-forming dwarf galaxies \citep[a median of $\approx 2.0$,][]{Cormier15} even if the flux losses due to poor sampling of extended \cii{} emission (surface brightness dimming, e.g., \citealt{Carniani20}) is taken into account. 
The origins of such a high \oiii{}-to-\cii{} ratio are still under debate \citep[e.g.,][]{Katz17, Katz19, Pallottini19, Lupi20, Arata20}, but the likely explanations include an intense and hard interstellar radiation field \citep{Harikane20, Witstok22}, which could make the bulk of the ISM highly ionized due to starbursts \citep{Ferrara19, Arata20, Vallini21}, and low C/O abundance ratios and high metal production efficiencies due to low-metallicity core-collapse supernovae with a top-heavy initial mass function \citep{Katz22}.

The rest-frame 90~$\micron$ continuum emission was also detected from \target{}, while the 160~$\micron$ continuum was not. This means that the spectral energy distribution (SED) at rest-frame 90--160 $\micron$ is close to the Rayleigh-Jeans regime with a high dust temperature like $T_{\rm dust} \sim 80$~K or even higher (\citealt{Bakx20, Sommovigo21}, see also \citealt{Sommovigo20}). This again implies that the ISM, restored in dense compact molecular clouds, is exposed to an intense and hard UV radiation field.  The inferred FIR luminosity and dust mass are $L_{\rm TIR} = 1 \times 10^{12}~L_\Sol$ and $M_{\rm dust} = 5 \times 10^5~M_\Sol$, respectively, if assuming $T_{\rm dust} = 80$~K and the dust emissivity index $\beta_{\rm dust} = 2$. Indeed, UV-to-FIR SED modeling\footnote{PANHIT \citep{Mawatari20} was used for the SED fits, in which an exponentially declining or rising SFR is assumed. For this galaxy, the SFR of the young stellar component can be approximated to be constant, since the best fitting $e$-folding timescale, $|\tau_{\rm SFH}| > 10$~Myr, is sufficiently longer than the stellar age. See \citet{Tamura19} for details.} of the stellar, nebular, and dust emission suggests the presence of young ($\approx 4$~Myr) stellar components with a high SFR of $\approx 60$--230~$M_\Sol$~yr$^{-1}$ as the origins of ionizing photons. This young component is inferred to have the stellar mass and metallicity of $\approx 2 \times 10^8~M_\Sol$ and $\approx 0.2~Z_\Sol$, respectively \citep{Tamura19}. Our dust evolution model \citep{Asano13}, however, does not reproduce the metallicity and dust mass in such a short duration of time, even though a high dust temperature is taken into account (see also e.g., \citealt{Michalowski15, Popping17, Vijayan19, Triani20, Dayal22} for theoretical attempts to explain the early evolution of dust mass).
\citet{Tamura19} found that an older (the age of $\sim 0.3$~Gyr), massive ($\sim 3 \times 10^9~M_\Sol$) stellar component which preexists by $z = 8.3$ can alleviate the tension without any conflict with the observed SED.  This underlying massive component is also suggested by a dynamical mass estimated from the \cii{} velocity field ($M_{\rm dyn} = (1.2 \pm 0.4) \times 10^{10}~M_\Sol$, \citealt{Bakx20}), which is much greater than the mass of the young stellar component.

Curiously, previous studies based on \emph{coarse resolution} ALMA imaging find that the dust and \cii{} emission is apparently co-spatial with the \oiii{} and UV emission on a galactic scale ($\sim 0\farcs 2$--$0\farcs 4$, corresponding to $\sim 1$--2~kpc), which is not consistent with the fact that the UV continuum is fairly blue (the UV slope of $\beta_{\rm UV} \approx 2$, \citealt{Tamura19}) with a small extinction. \target{} has three individual clumps of young stars detected in the \hst{}/WFC3 image. The dust emission peaks in between the eastern and central clumps, but the overall distribution is similar to the UV and \oiii{} distributions. In general, the presence of dust is expected to enhance the amount of molecular gas, because the surfaces of dust grains are the formation site of H$_2$ molecules \citep[e.g.,][]{Hirashita02}. This means that the dust emission may trace molecular clouds within this galaxy. Therefore, a possible explanation for the co-spatial distribution is patchy/porous geometry of dusty molecular clouds and highly ionized gas on sub-kpc scales. This picture is all suggested by state-of-the-art hydrodynamic simulations of galaxy formation \citep[e.g.,][]{Arata19, Arata20, Liang19, Pallottini22}, while the current ALMA imaging is not sufficient in angular resolution to depict the internal structure of the galaxy.

Here we present the results of $\approx 300$~pc resolution imaging of this $z = 8.312$ LBG, \target{}, in the \oiii{} 88~$\micron$ line and the rest-frame 90~$\micron$ (i.e., the observer-frame 850~$\micron$) dust continuum, in order to directly compare the sub-kpc scale distributions of young massive stars (UV), highly ionized gas (\oiii{}), and molecular clouds (dust). The effective resolution for the dust continuum and \oiii{} images reaches 340~pc $\times$ 390~pc and 290~pc $\times$ 360~pc, respectively (see Section~\ref{sec:delensing}), which is the first and furthest imaging resolving multi-phase ISM in a galaxy found in the heart of the reionization era, when the age of the Universe was only 600~Myr.

The structure of this paper is as follows. Section~\ref{sec:observations} describes ALMA observations and ancillary data. The results of the ALMA observations are presented in Section~\ref{sec:results}. In Sections~\ref{sec:spatial} and \ref{sec:discussions}, we will discuss the spatial distribution of stars, ionized gas and dust on a 300~pc scale. Finally, Section~\ref{sec:conclusions} concludes the paper.

Throughout this paper we assume a cosmology with $H_0=70$~\kms{}~Mpc$^{-1}$, $\Omega_{\rm m,0} = 0.3$, and $\Omega_{\Lambda,0}=0.7$. The angular size of $1\farcs 0$ corresponds to 4.7~kpc at $z=8.31$.


\section{Observations and Ancillary Data} \label{sec:observations}

\begin{deluxetable*}{ccccccc}
\tablecaption{Log of the ALMA observations\label{tab:alma_log}}
\tablewidth{0pt}
\tablehead{
\colhead{Program ID} & \colhead{UT start date} & \colhead{$L_{\rm basel}$} & \colhead{$N_{\rm ant}$} & 
    \colhead{Tuning} & \colhead{$t_{\rm int}$} & \colhead{PWV} \\
\colhead{} & \colhead{(YYYY-MM-DD)} & \colhead{(m)} & \colhead{} &
    \colhead{} & \colhead{(min)} & \colhead{(mm)}
}
\decimalcolnumbers
\startdata
2016.1.00117.S$^{\dagger}$ & 2016-10-25 & 19--1399 & 43 & T2 & 32.76 & 0.62 \\
 & 2016-10-26 & 19--1184 & 46 & T2 & 32.76 & 0.30 \\
 & 2016-10-28 & 19--1124 & 39 & T3 & 32.30 & 0.35 \\
 & 2016-10-29 & 19--1124 & 41 & T1 & 33.77 & 1.27 \\
 & 2016-10-30 & 19--1124 & 39 & T3 & 38.30 & 0.93 \\
 & 2016-10-30 & 19--1124 & 40 & T1 & 33.77 & 0.78 \\
 & 2016-11-02 & 19--1124 & 40 & T4 & 30.23 & 0.64 \\
 & 2016-11-02 & 19--1124 & 40 & T4 & 30.23 & 0.97 \\
 & 2016-12-17 & 15--460  & 44 & T1 & 33.77 & 0.90 \\
 & 2016-12-18 & 15--492  & 47 & T1 & 33.77 & 1.29 \\
 & 2017-04-28 & 15--460  & 39 & T3 & 38.30 & 0.72 \\
 & 2017-07-03 & 21--2647 & 40 & T4 & 30.23 & 0.24 \\
 & 2017-07-04 & 21--2647 & 40 & T4 & 30.23 & 0.41 \\
 \hline
2017.1.00225.S & 2018-08-12 & 15--483 & 43 & T3$^\ddag$ & 44.08 & 0.51 \\
 & 2018-08-12 & 15--440 & 43 & T3$^\ddag$ & 44.07 & 0.74 \\
 & 2018-08-19 & 15--440 & 43 & T3$^\ddag$ & 26.15 & 0.68 \\
 & 2018-12-24 & 15--500 & 43 & T4 & 43.58 & 0.41 \\
 & 2018-12-25 & 15--500 & 43 & T4 & 43.60 & 0.44 \\
 & 2018-12-25 & 15--500 & 43 & T4 & 43.55 & 0.49 \\
 \hline
2017.1.00486.S & 2018-05-22 & 15--314 & 44 & Ta & 35.40 & $\cdots^\flat$ \\
 & 2018-05-31 & 15--314 & 45 & Ta & 35.40 & $\cdots^\flat$ \\
 & 2018-06-21 & 15--313 & 43 & Tb & 37.93 & $\cdots^\flat$ \\
 & 2018-06-21 & 15--313 & 43 & Tb & 37.93 & $\cdots^\flat$ \\
 & 2018-05-20 & 15--314 & 44 & Tc & 30.35 & $\cdots^\flat$ \\
 & 2018-05-20 & 15--314 & 44 & Tc & 30.35 & $\cdots^\flat$ \\
 \hline
2018.1.01241.S
 & 2018-10-09 & 15--2516 & 43 & T4$^\ddag$ & 48.15 & 0.84 \\
 & 2018-10-10 & 15--2516 & 43 & T4$^\ddag$ & 48.17 & 0.60 \\
 & 2018-10-10 & 15--2516 & 43 & T4$^\ddag$ & 48.15 & 0.55 \\
 & 2018-10-14 & 15--2516 & 43 & T4 & 48.15 & 0.56 \\
 & 2018-10-15 & 15--2516 & 43 & T4 & 48.17 & 0.60 \\
 & 2018-10-15 & 15--2516 & 43 & T4 & 48.17 & 0.58 \\
 & 2018-10-18 & 15--2516 & 43 & T4 & 48.17 & 0.31 \\
 & 2019-08-14 & 41--3143 & 43 & T4 & 48.15 & 0.49 \\
 & 2019-08-20 & 41--3637 & 43 & T4 & 48.17 & 0.37 \\
 & 2019-08-24 & 41--3396 & 43 & T4 & 48.13 & 0.54 \\
 & 2019-08-24 & 41--3396 & 43 & T4 & 48.15 & 0.64 \\
 & 2019-08-24 & 41--3396 & 43 & T4$^\flat$ & 48.12 & 0.55 \\
 & 2019-08-26 & 41--3637 & 43 & T4 & 48.15 & 0.66 \\
 & 2019-08-28 & 41--3637 & 43 & T4 & 48.17 & 0.73 \\
\enddata
\tablecomments{(1) Program identifier, (2) start time of the observation in UTC, (3) baseline length of the used array configuration, (4) the number of the 12-m antennas, (5) tuning identifier, (6) integration time, and (7) precipitable water vapor column. The tuning identifiers include T1, T2, T3, T4, Ta, Tb, and Tc with the local oscillator frequencies of 347.80, 351.40, 355.00, 358.60, 347.62, 354.61, and 358.08~GHz, respectively. Frequency division mode (FDM) of the correlators was used for tunings T1 to T4, whereas the time division mode (TDM) for tuning Ta to Tc. $^\dagger$Reported by \citet{Tamura19}. $^{\ddag}$Semi-passed data. $^{\flat}$Missing in a quality assurance (QA2) report.}
\end{deluxetable*}

\subsection{ALMA observations and calibration}

Table~\ref{tab:alma_log} summarizes the ALMA Band~7 observations. The observations were performed over Cycles 4 to 6 from 2016 to 2019 (project IDs: 2016.1.00117S, 2017.1.00225.S, 2017.1.00486.S, 2018.1.01241.S), part of which was already reported by \citet{Tamura19}. Frequency setups are also summarized in Table~\ref{tab:alma_log}. The spectral windows (SPWs) were originally configured to cover as wide a frequency band as possible to search for \oiii{} to identify its spectroscopic redshift. Here only two SPWs, T4/SPW2 and Tc/SPW3, cover the \oiii{}, while the remaining SPWs can only be used for continuum imaging. Total on-source times for continuum and \oiii{} are 27.52~hr and 18.89~hr, respectively. This is 3.8$\times$ and 9.4$\times$ improvement in continuum and line integration time \citep[previously 7.26~hr and 2.02~hr in][]{Tamura19}, respectively. Calibration for most of the datasets was performed in the standard manner using CASA \citep{McMullin07} pipelines provided by the observatory. For `semi-passed' datasets, which are partially quality-assured by the observatory, we used the CASA (version 5.6.1) to generate a pipeline and processed them for calibration. We carefully flagged corrupt data we found among the pipeline-processed semi-passed datasets.


\begin{figure}[t!]
\includegraphics[width=0.48\textwidth]{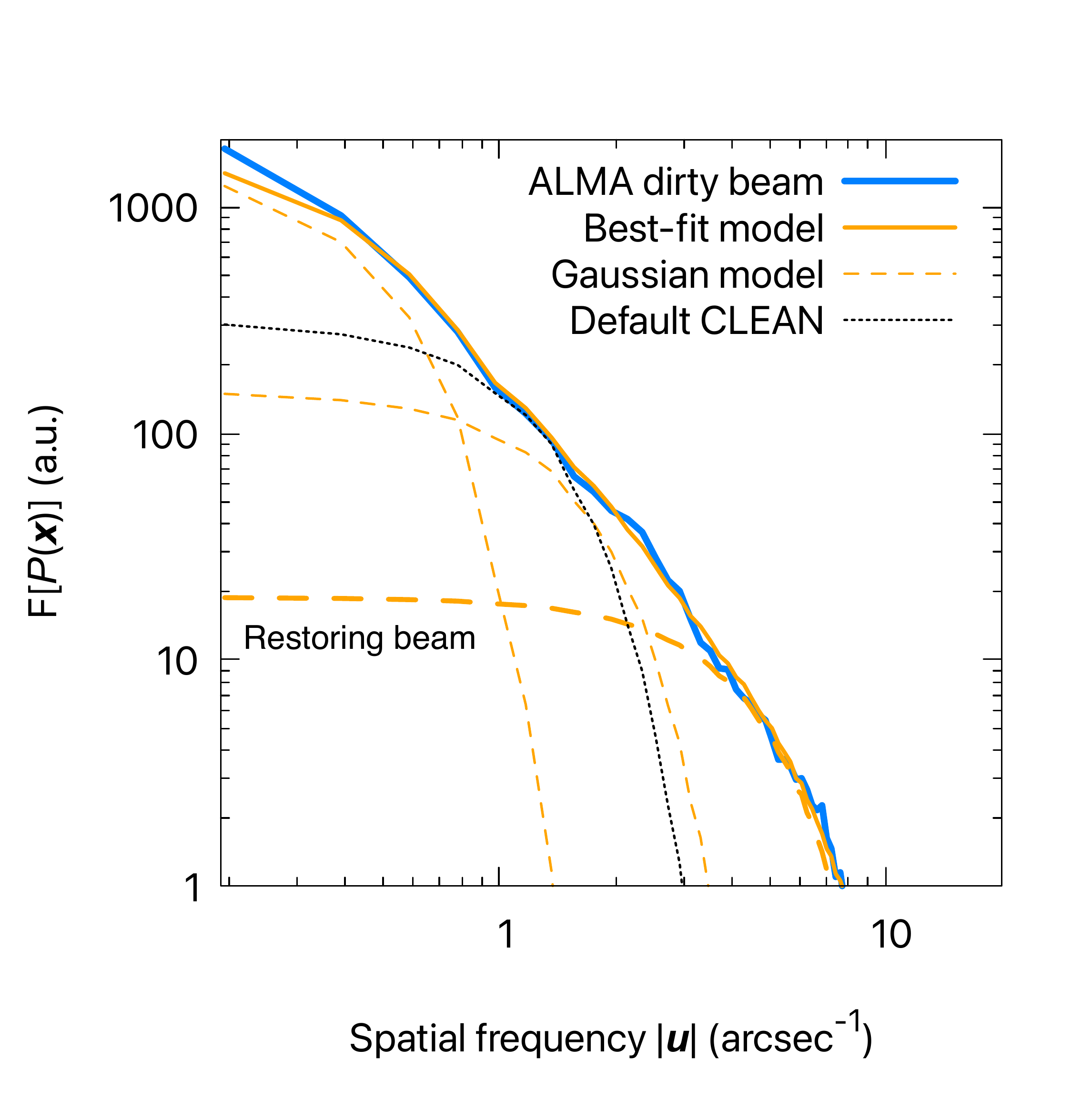}
\caption{
The radial profile of the Fourier transform of the point spread function (PSF, or a synthesized beam) obtained at the 850~$\micron$ continuum observations (blue solid curve). The ($u,\,v$)-coverage of our ALMA observations is not circularly symmetric but rather has an elliptical shape with the minor-to-major axis ratio of 0.8. For a display purpose, we compensate this elongation by stretching the Fourier transform by 1/0.8 in the direction of the minor axis when the radial profile is made.  The solid orange curve shows the best-fitting model which is composed of 3 Gaussian functions with different widths (dashed curves). The Gaussian with the highest frequency (i.e., the thick dashed curve) is used for the restoring beam. 
The dotted curve shows the radial profile of the Fourier transform of the CLEAN beam that the default CLEAN algorithm uses. \label{fig:psf}}
\end{figure}



\begin{figure*}[ht!]
\includegraphics[width=1.00\textwidth]{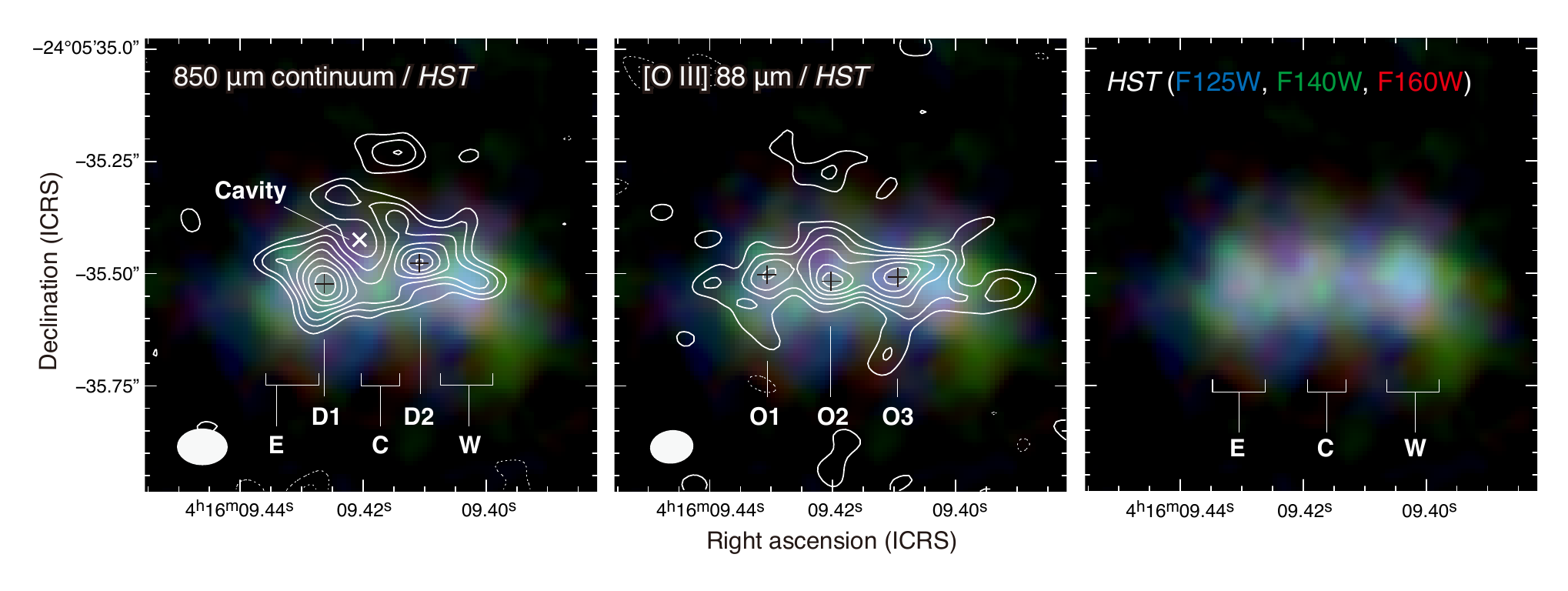}
\caption{The 850~$\micron$ continuum (left), \oiii{} 88~$\micron$ (middle), and the rest-frame UV (right) images obtained for \target{}. The contours are $(2, 3, 4, \cdots) \times \sigma$, where $\sigma = 3.7$~$\mu$Jy~beam$^{-1}$ (dust) and 11~mJy~beam$^{-1}$~\kms{} (\oiii{}). The negative contours are shown with the dotted lines. The ellipse at the bottom-left corner of the left or middle panel represents the beam FWHM. The background shows a composite color image comprising the F125W (blue), F140W (green), and F160W (red) filters of \hst{}/WFC3. The black crosses mark the positions of dust and \oiii{} peaks (D1, D2, O1, O2, and O3, respectively). The white \textsf{x} symbol on the left panel shows the position of the cavity. `E', `C', and `W' denote the positions of the eastern, central, and western stellar clumps seen in the \hst{}/WFC3 image, respectively.  \label{fig:alma_hubble}}
\end{figure*}


\subsection{ALMA imaging and beam restoration}

We use the CASA task \texttt{tclean} to Fourier-transform the visibility data. 
We employed the natural weighting  to maximize the point-source sensitivity for the continuum and \oiii{} imaging. The continuum image is produced from all of the spectral windows except for the frequency range of 364.1--364.7~GHz, where \oiii{} emission is observed. The \oiii{} data is continuum-subtracted before imaging. The \oiii{} cube is re-sampled at the frequency resolution of 15.625~MHz, which corresponds to a velocity resolution of 12.85~\kms{} at the line frequency. The standard CLEAN deconvolution \citep{Hogbom74} is made down to twice the r.m.s.\ level of each of the continuum and \oiii{} images.

Figure~\ref{fig:psf} shows the radial profile of the 2-dimensional Fourier transform of the synthesized beam or point-spread function (PSF) obtained for the 850~$\micron$ continuum observations.  Since the resulting image comprises multiple observing runs where different array configurations were employed (Table~\ref{tab:alma_log}), the main-lobe of the PSF has multiple spatial frequency components, which are well described by a linear combination of $\approx 3$ Gaussian functions. This means that the PSF is composed of a sharp peak with an extended tail around it.

The tail, however, substantially degrades the effective angular resolution of the deconvolved image. 
In the deconvolution process, as implemented in \texttt{tclean}, it forcefully fits the PSF to a single 2-dimensional Gaussian and assumes this Gaussian for a restoring beam or a so-called CLEAN beam (see the dotted curve in Figure~\ref{fig:psf}).
The CLEAN algorithm \citep{Hogbom74} includes deconvolution and CLEAN-beam restoration, which are mutually independent processes. In general the latter is done in order to compensate for the unrealistic nature of CLEAN-based component images that include delta functions. Choice for the CLEAN beam is somewhat arbitrary, and desirable characteristics of a CLEAN beam are (1) it is free from sidelobes, and (2) its Fourier transform should be constant inside the sampled region of the $(u,\,v)$ plane and rapidly fall to a low level outside it \citep[\S~11.1,][]{Thompson17}.

As seen in Figure~\ref{fig:psf}, the PSF is well fitted with a combination of 3 concentric 2-dimensional Gaussian components with a range of full-widths at half maximum (FWHMs). For the continuum (\oiii{} line) PSF, we find that the best-fitting Gaussian which is responsible for the sharpest resolution has FWHM = $81.1 \times 112.2$~mas with a position angle (PA) of 89.4$\arcdeg$ ($73.9 \times 96.1$~mas with PA = 93.8$\arcdeg$).
Here we employ the sharpest one for the CLEAN beam (see the thick dashed curve in Figure~\ref{fig:psf}), which always meets the characteristics required for the CLEAN beam. 
The beam restoration was done using the \texttt{tclean} parameter, \texttt{restoringbeam}, and was added to the CLEAN residual image. The resulting r.m.s.\ noise levels obtained for the 850~$\micron$ continuum image and the \oiii{} cube with a 15.625~MHz resolution are 3.7~$\mu$Jy~beam$^{-1}$ and 0.13~mJy~beam$^{-1}$, respectively.
\footnote{The noise level could increase by $\sim 30\%$ if the default clean beam is used for beam restoration. This is probably because the beam restoration adds beam-sized Gaussians at the positions where $>2\sigma$ noise components are identified as the clean component, leaving large-scale fluctuations to the resulting image.}

One should note that the CLEAN deconvolution can cause incorrect flux scaling when the PSF is largely different in solid angle from the CLEAN beam, regardless of whether the CLEAN beam is selected by the default choice or is chosen optimally like what is done in this study. This is due to a mismatch between units of the restored and residual images, which are Jy per CLEAN beam and per PSF, respectively. The CLEAN algorithm simply adds the two images without treating their units appropriately to make a final product image, which is in units of Jy per CLEAN beam. In that case, a low-level flux embedded only in the residual image might be overestimated in the final image if the CLEAN beam is much smaller in solid angle than the PSF.  This is called the CLEAN bias and originally pointed out by \citet{Jorsater95} \citep[see also][]{Czekala21}. In our case, the brightness of a low-level extended emission below the CLEAN threshold (i.e., $2\sigma$ for our study) might be overestimated and should be treated with caution. We find in our new continuum image that the source flux density fallen within the $2\sigma$ contour is $\approx 149~\mu$Jy, in good agreement with the previous study ($S_\mathrm{850~\mu m} = 137 \pm 26~\mu$Jy, \citealt{Tamura19}), where little CLEAN bias was expected. The flux density within the $1\sigma$ contour is, however, $\approx 194~\mu$Jy, $\sim 30\%$ larger than the previous value, implying the presence of the CLEAN bias in the lower brightness level.

\subsection{Hubble Images and Astrometric Considerations}

The astrometric error needs to be corrected before comparing with \hubble{} imaging, as the reference frames to which the ALMA and \hst{} images are registered are different,  For this sake, we use four {\it Gaia} sources \citep{Gaia16, Gaia20} found in the \hst{}/WFC3 F160W image to correct an astrometric offset between the ALMA and \hst{} images. The {\it Gaia} sources include two stars and two galaxies and are registered to the ICRS coordinate system, on which ALMA phase calibration also relies. The correction was done in the same manner presented by \citet{Tamura19}. The residual offsets found for the stars and galaxies after the correction were almost negligible. The average offset from ICRS is ($\Delta$RA, $\Delta$Dec) = ($4 \pm 7$ mas, $-4 \pm 12$ mas), where the uncertainties are the standard deviations. The individual offsets are, however, $\approx 20$~mas (1/5 of the ALMA beam FWHM) for the most erroneous case, and hereafter we take this as the systematic uncertainty of astrometric correction between the ALMA and \hst{} images.
Note that ALMA itself has a systematic astrometric uncertainty due to baseline calibration errors, which is approximately 3~mas (ALMA Technical Handbook\footnote{https://almascience.nrao.edu/documents-and-tools/documents-and-tools/}).


\begin{figure*}[ht!]
\includegraphics[width=\textwidth]{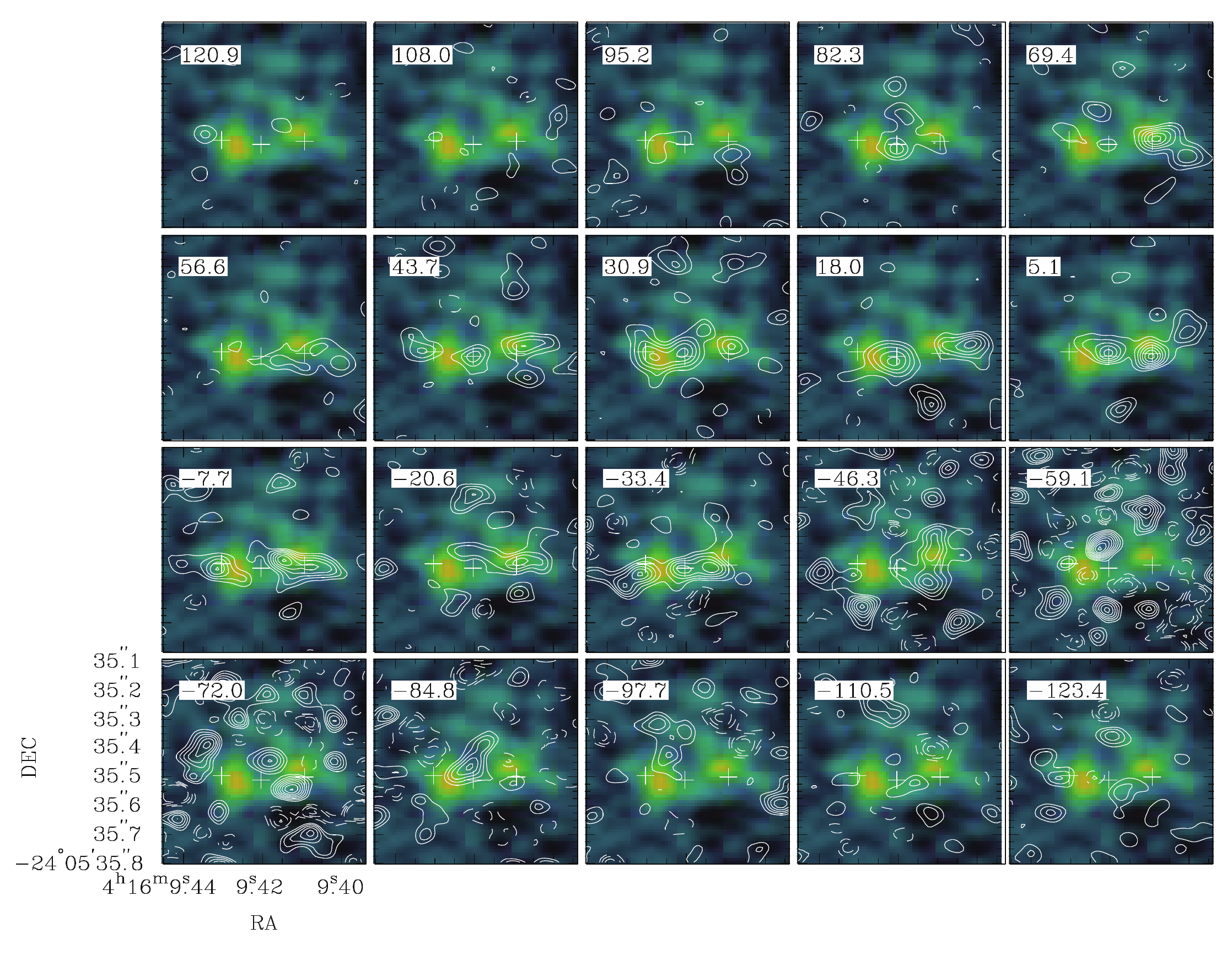}
\caption{The \oiii{} 88~$\micron$ channel maps of \target{} (contours) overlaid on the dust continuum image (background image). The contours are drawn at $(2,\, 3,\, 4,\, \cdots) \times \sigma$, where we adopt $\sigma = 0.13$~mJy~beam$^{-1}$ for all panels.  The negative values are indicated by the dashed contours. The crosses mark the positions of the \oiii{} peaks, O1, O2, and O3. The velocity offset (in units of \kms) is given at the top-left corner of each panel. The systemic velocity is set to 364.350~GHz, and the channel separation is 12.85~\kms{}. Note that the channels at $-59.1$ and $-72.0$ \kms{} are contaminated by the atmospheric absorption line, and the r.m.s.\ values at these channels are approximately twice as large as the $1\sigma$ value given above. \label{fig:channel_map}}
\end{figure*}

\begin{figure*}[ht!]
\includegraphics[width=\textwidth]{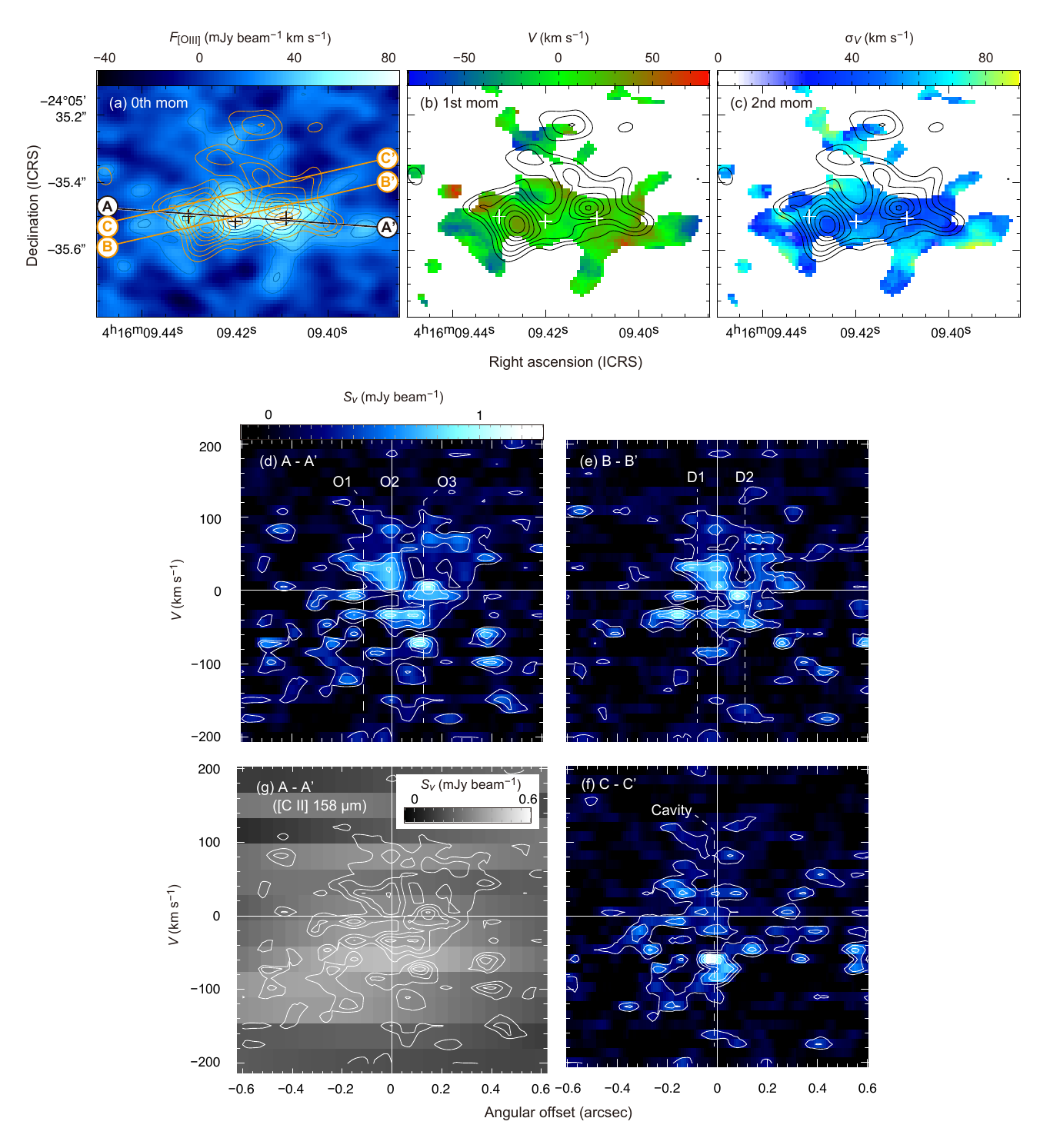}
\caption{{\it Top}: The \oiii{} 88~$\micron$ moment images of \target{}. (a) The 0-th moment image (integrated intensity image) of the \oiii{} 88~$\micron$ map, overlaid with the dust continuum image with the orange contours. 
(b) The first moment image (velocity field).
(c) The second moment image (velocity dispersion). The contours show the dust continuum emission. The contour levels are the same as those displayed in Figure~\ref{fig:alma_hubble}. The crosses show the \oiii{} peaks. 
{\it Bottom}: The position--velocity diagrams (PVDs) of \target{} as seen in \oiii{} 88~$\micron$ and \cii{} 158~$\micron$. (d) The PVD extracted along the black solid line denoted as A--A' on panel (a). (e, f) The same as (d), but along B--B' and C--C' penetrating the dust peaks and cavity, respectively (the orange solid lines on panel (a)). (g) The same as (d), but the background image shows the \cii{} emission \citep{Bakx20}.  In panels (d) to (g), the \oiii{} contours are drawn at (1.5, 3, 4.5, $\cdots$) $\times 0.13$~mJy~beam$^{-1}$. 
Note that the atmospheric absorption line is contaminating in the velocity range of $\approx -60$ to $-70$~\kms{}. In panel (f), a bright $-59.1$~\kms{} knot at the cavity position is seen, but the significance is not high ($\sim 4\sigma$) on the $-59.1$~\kms{} channel map.
\label{fig:moment_map}}
\end{figure*}


\section{Results} \label{sec:results}

Figure~\ref{fig:alma_hubble} shows the 850~$\micron$ continuum and \oiii{} integrated images overlaid on the \hst{} NIR image.
The continuum image clearly exhibits two clumps (D1 and D2), which are apparently bridging over the 3 UV clumps (E, C, and W). The offsets between the peak positions of the dust and UV clumps are significant, given the astrometric accuracy better than $\approx 20$~mas, implying the presence of dark lanes separating an intrinsically continuous UV star cluster or the three young star clusters forming close to the surface of the two dusty molecular clouds that are seen in the 850~$\micron$ continuum. 
The latter might be the case for the clumps D1 and E, since the D1--E offset is relatively small.
The sizes of the dust and UV clumps are typically $\lesssim 0\farcs 2$. We will revisit the intrinsic physical scales of those clumps in \S~\ref{sec:spatial}. The dust emission shows a fairly asymmetric distribution in the direction perpendicular to the stellar extent. The southern part of dust emission drops off steeply, whereas the emission extends toward the north, with a galaxy-wide clumpy tail. This feature is very similar to that found in CO~(2--1) of a nearby dwarf, Haro~2, where starburst-driven outflows and a superbubble are found \citep{Beck20}.

The dust continuum image of \target{} shows several off-center depressions embedded in the clumpy tail. Although many of them are low-significance and we cannot tell if they are real, the one located at the interface between the clumpy tail and the central UV clump, C, is more prominent (see the cross symbol in Figure~\ref{fig:alma_hubble} left). Its diameter is $\approx 150$~mas and is barely resolved with our $81 \times 112$~mas beam.  We will discuss in Section~\ref{sec:superbubble} whether a kpc-scale cavity can be produced as a result of stellar feedback.

The \oiii{} emission is more elongated in the east--west direction than the dust. It has three peaks, O1, O2 and O3, neither of which coincides with D1 or D2 (see also Figures~\ref{fig:channel_map} and \ref{fig:moment_map}a). The emission basically traces the UV continuum emission. This is naturally expected because \oiii{} traces high ionizing flux incident from young hot stars. It seems that a part of \oiii{} is also associated with the dust component, suggesting the presence of obscured star-formation or a chance association along the line of sight. Unlike the dust emission, \oiii{} traces filamentary structures. It also shows several spur-like structures extending from the O1 and O3 peaks, although the significance is not high (up to $3\sigma$).
There is an offset between O3 and W, and little \oiii{} emission is associated with W, whereas the clump W is as blue in the UV as E and C (see the middle and right panels of Figure~\ref{fig:alma_hubble}). This could be due to a low \oiii{} emissivity of W because of dense ionized gas, which de-excites the \oiii{} 88~\micron{} transition. Another possibility is that stellar feedback could have removed a significant fraction of the gas from this region. The western `spurs' in the \oiii{} distribution could possibly support this blow-out scenario. Another option could be that W is actually subdominant in producing ${\rm O}^+ \rightarrow {\rm O}^{++}$ ionizing photons compared to E and C, and just appears bright and blue in the \hst{} image because of less dust reddening. \jwst{}/NIRSpec integral field spectroscopy of the rest-frame optical \oiii{}/H$\beta$ lines and NIRCam imaging of the stellar continuum will address the gas properties and possible reddening of the UV clumps.

Figure~\ref{fig:channel_map} shows the channel maps in step of 15.625~MHz in observing frequency or 12.85~\kms{} in Doppler velocity, which reveals a very complicated structure of \oiii{}. We find that the \oiii{} peaks O1 to O3, indicated by the crosses, have multiple velocity components, while each component only has a relatively small velocity dispersion ($\sim 10$~\kms{}). One should note that the atmospheric absorption line is contaminating at $V \approx -70$ to $-50$~\kms{}.
Another caveat is that the presence of spatially extended emission is not ruled out. Interferometric imaging may fail to reproduce an extended emission and discrete it into clumpy structures instead, when the significance of the emission is not high \citep{Gullberg18}.  If such clumpy artifacts appear independently among spectral channels, a single broad emission line may also be split into multiple velocity components, mimicking a complicated velocity structure. 
A roughly twice deeper sensitivity would be necessary to confirm the velocity structure, although this is not easy for ALMA to achieve because it requires $\sim 4\times$ more integration time than the current one (18.9~hr). The planed ALMA wideband sensitivity upgrade \citep[WSU,][]{Carpenter22} will improve the line sensitivity slightly (20--50\%), which will make the deeper integration feasible.

Figures~\ref{fig:moment_map}b and \ref{fig:moment_map}c show the velocity-field and velocity-dispersion images where noisy channels at $364.430 \pm 0.015$~GHz and low significance pixels ($< 2\sigma$ of the integrated intensity image) are removed before making the moment images. The integrated intensity image is also shown in Figure~\ref{fig:moment_map}a for comparison. The velocity field is turbulent and shows no clear global gradient. The intensity-weighted global velocity dispersion\footnote{$\langle \sigma_{V} \rangle_{\rm w} \equiv \sum_{I > 2\sigma}{\sigma_{V}I}/\sum_{I > 2\sigma}{I}$, where $\sigma_{V}$ and $I$ are the velocity dispersion and the integrated intensity of \oiii{}, respectively.} averaged over the region where the \oiii{} intensity is detected at $> 2\sigma$ is estimated to be $\langle \sigma_{V} \rangle_{\rm w} = 54$~\kms{}. One should note that a small velocity gradient which is comparable to the global velocity dispersion might exist on a large spatial scale in the east-west direction, which is consistent with that found in \cii{} \citep{Bakx20}. 
This means an apparently small rotation-to-dispersion ratio of $V_{\rm rot}/\sigma_{V} \sim 1$, suggesting either a dispersion-dominated system and/or a rotation-dominated disk with a very small inclination angle (i.e., in a face-on view). 
If the latter is the case, the global velocity dispersion is consistent with those predicted for a face-on rotating disk with a stellar mass of $M \sim 10^9 M_\odot$ at $z \sim 6$--8 \citep{Kohandel20}.
The velocity dispersion around the \oiii{} brightness peaks, O2 and O3, is relatively small ($\sigma_{V} \sim 40$--50~\kms{}). This is likely because a few velocity components of \oiii{}-emitting gas dominate the brightness toward each of the O2 and O3 peaks. In contrast, the outskirts, especially the northern tail surrounding the dust cavity, show a large dispersion ($\sigma_{V} \sim 80$~\kms{}). This is probably because multiple velocity components with similar brightness contribute to the apparent broadening of the emission line, or a high turbulent velocity due to low gas density in the northern tail, 
although relatively low significance of emission toward the outskirts would mimic the velocity dispersion.

Figure~\ref{fig:moment_map} (bottom) shows a set of the position--velocity diagrams (PVDs) extracted along the lines crossing the three \oiii{} peaks (A--A'; see the line in Figure~\ref{fig:moment_map}a), the two dust peaks (B--B'), and the cavity (C--C'). In Figure~\ref{fig:moment_map}d we show the PVD crossing the \oiii{} peaks. As suggested from the channel maps, we find that the emission in any line of sight has multiple velocity components, each of which has a series of relatively narrow velocity components ($\Delta V \lesssim 20$~km/s) on top of a broader one ($\sim 100$--150~km/s). This is similar to earlier results from hydrodynamic simulations predicting the FIR line profiles \citep[e.g.,][]{Vallini13, Pallottini19, Kohandel19}, while the ALMA result suggests that the system is comprised of a collection of more numerous \hii{} regions with random bulk motion. This can be related to the prediction from numerical simulations that the \oiii{} emission traces individual ionized regions that are localized close to young massive stars \citep{Pallottini19, Arata20}, leading to a patchy and clumpy structure of \oiii{} in the predicted integrated intensity and velocity field maps compared with the \cii{} maps \citep{Katz19}.

Figure~\ref{fig:moment_map}e shows the PVD crossing D1 and D2 (B--B'). The \oiii{} emission is seen even along the sightlines of D1 and D2, whereas there are several depressions at $V \approx -20$ and $-60$~\kms{} for D1 and 30~\kms{} for D2. The depressions might be the regions where dusty molecular clouds are dominant. We also find the \oiii{} emission in the blueshifted part ($V \lesssim -50$~\kms{}) fainter than that seen in the PVD A--A'. This could be due to a low ionization fraction of the ISM in the blueshifted components.  
In Figure~\ref{fig:moment_map}g, we show the PVD of the \cii{} 158~$\micron$ line, which generally traces neutral ISM, although the resolution is coarser ($0\farcs46 \times 0\farcs64$) than \oiii{}.  The \cii{} emission is more prominent in the blueshifted part and shows a large-scale velocity gradient, which is not significant in \oiii{}. Indeed, the systemic velocity of \cii{} is blueshifted \citep[$z=8.3113 \pm 0.0003$,][]{Bakx20} with respect to the \oiii{} redshift ($z = 8.3125$).  This suggests a velocity-space segregation of multi-phase ISM in \target{}.
We note that the large-scale velocity gradient seen in \cii{} is not likely due to the difference between the \oiii{} and \cii{} angular resolutions, because our previous imaging of \oiii{} at a similar resolution ($0\farcs 38 \times 0\farcs 36$, \citealt{Tamura19}) showed no velocity gradient in \oiii{}. The large velocity offset can be due to an intrinsic difference in the spatial and/or velocity distribution between \oiii{} and \cii{}, which is also predicted by recent hydrodynamic simulations where the smoother and more extended distribution of \cii{} than \oiii{} is evident \citep{Katz19, Pallottini19, Arata20}.

The PVD C--C' (Figure~\ref{fig:moment_map}f) shows the velocity structure around the dust cavity. As revealed in the second moment image (Figure~\ref{fig:moment_map}c), the \oiii{} emission is scattered in a broad velocity range ($\sigma_{V}\sim 80$~\kms{} or the full line width of $\Delta V \sim 200$~\kms{}).  The emission is fragmented into multiple components, although we find no broad cavity in the velocity domain that is suggestive of a simple expanding shell. This velocity structure is similar to that found for a 100-pc cavity of 30~Doradus in the Large Magellanic Cloud, where broadened optical H$\alpha$, \oiii{}~5007~\AA{}, and \textsc{[N~ii]} lines with multiple velocity components are identified across the cavity embedded in the nebula \citep{Melnick99, Torres-Flores13, Castro18}
although its size is much smaller ($\sim 1/3$) than the cavity found in \target{}.


\section{Spatial distribution of dust, ionized gas, and young stars} \label{sec:spatial}

In Section~\ref{sec:results}, we revealed the clumpy and filamentary morphology of the dust and \oiii{} emission. It also shows that the velocity field traced by the \oiii{} emission is dispersion-dominated.  The different spatial distributions of dust and \oiii{} suggest that we resolve the multi-phase ISM in \target{}, which can provide us with a spatial scale that is a key to characterizing the physical properties of ISM.  In this section we investigate the spatial distributions of dust, ionized gas, and the young stars while correcting for gravitational lensing magnification.


\begin{figure*}[t!]
\includegraphics[width=\textwidth]{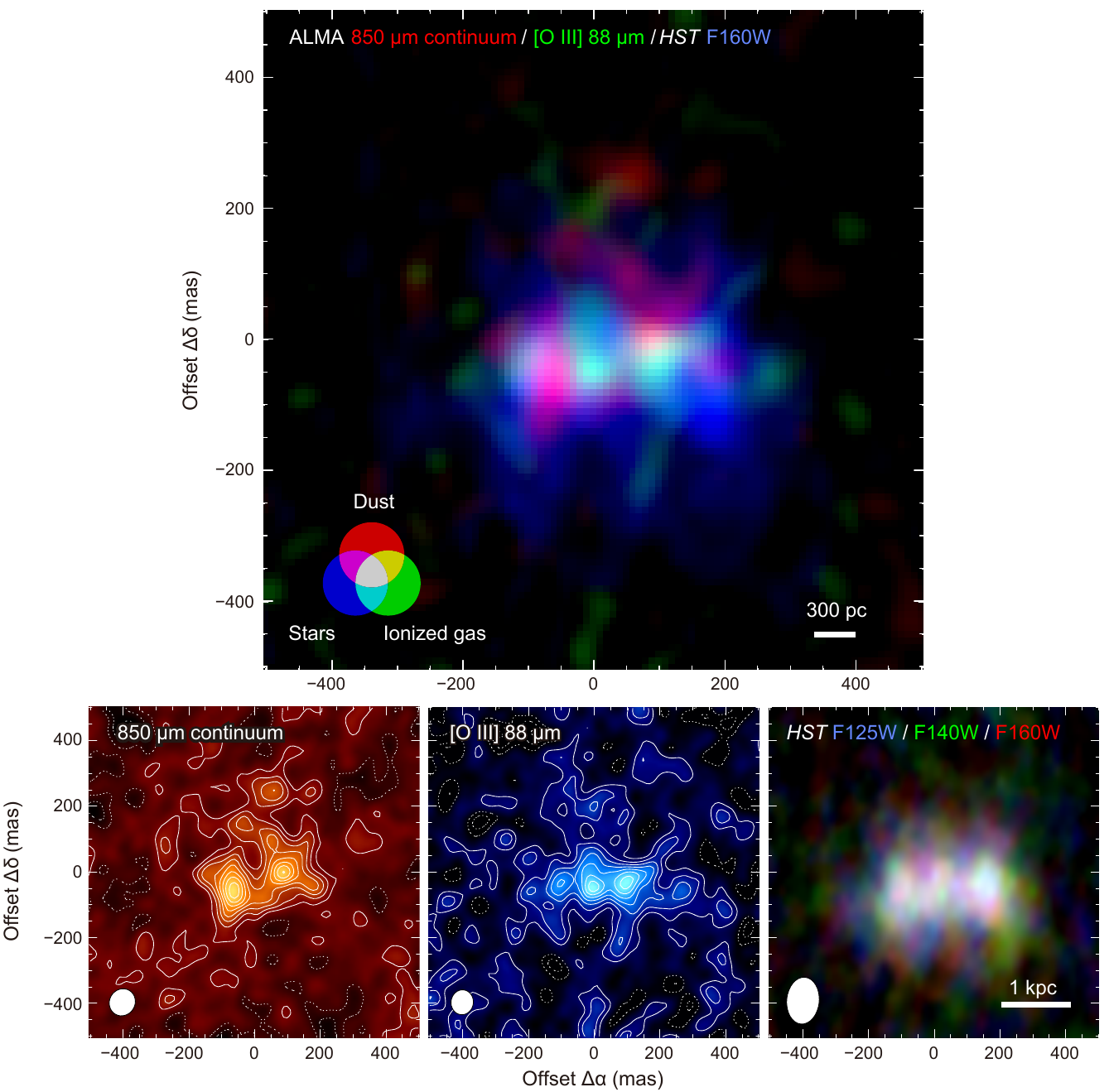}
\caption{
    The delensed 3-color image of \target{} (top) composed of dust (bottom-left), \oiii{} 88~$\micron$ (bottom-center), and the rest-frame UV continuum (bottom-right). The effective angular resolution is represented by the filled ellipse at the bottom-left corner of each of the bottom panels. The scale bar on the top and bottom-right panel indicates the 300~pc and 1~kpc scale on the source plane, respectively.  The contours are drawn at $\pm(1,2,3,...)\times \sigma$, where $\sigma = 3.7$~$\mu$Jy~beam$^{-1}$ and 11~mJy~beam$^{-1}$~\kms{} for the dust and \oiii{} images, respectively. \label{fig:demagnified}}
\end{figure*}


\subsection{Correction for gravitational lensing}\label{sec:delensing}

As \target{} is gravitationally magnified by the foreground cluster, it needs to be corrected for the lensing magnification in order for quantitative analyses on true distributions of stars, ionized gas, and dust. 
To correct for magnified (i.e., apparent) ALMA and \hubble{} images, we employ a lensing model of \citet{Kawamata16} produced using \textsc{Glafic} \citep{Oguri10} among several available models because of its relatively better performance 
in reproducing mass distributions of simulated clusters of galaxies \citep{Meneghetti17} and in predicting the mass distribution of the MACS~J0416.1$-$2403 cluster based on the maximum number of lensed images \citep{Priewe17}.
The choice, however, does not dramatically change the result, since the \target{} is located sufficiently away from the critical lines produced by the cluster.  We find no strong lensing sources close to \target{} and take into account the local convergence ($\kappa$) and shear ($\gamma_1$, $\gamma_2$) parameters to make a delensed image. The typical values for them are $(\kappa,\,\gamma_1,\,\gamma_2) = (0.13520,\, 0.15505,\, -0.02532)$ at the position and redshift of \target{}.%
\footnote{The typical values for $\kappa$, $\gamma = (\gamma_1^2 + \gamma_2^2)^{1/2}$, and $\mu_{\rm g}$ from 9 available lens models are 
$\kappa = 0.206 \pm 0.089$, $\gamma = 0.176 \pm 0.041$, and $\mu_{\rm g} = 1.57 \pm 0.36$. The variations for $\gamma$ and $\mu_{\rm g}$ are both $\approx 23\%$, which means that the delensed size scale and aspect ratio could change $\sim 10\%$ and $\sim 20\%$, respectively.}

The effective spatial resolutions of the demagnified images are estimated as follows.  The inversion matrix is expressed as
\begin{eqnarray}
{\bf A} = 
(1-\kappa){\bf I} -
\left(
    \begin{array}{ccc}
         \gamma_1 & -\gamma_2 \\
        -\gamma_2 & -\gamma_1
    \end{array}
\right)\,,
\end{eqnarray}
where ${\bf I}$ is the identity matrix.  The demagnified beam shape (i.e., an ellipse) representing the effective angular resolution on the source plane is expressed by the following quadratic equation,
\begin{eqnarray}
{\bf x}^{\rm T}({\bf A}^{-1})^{\rm T}{\bf B}{\bf A}^{-1}{\bf x}=1 \,,
\end{eqnarray}
where ${\bf x}$ is the angular coordinate on the sky, and $\bf B$ is the matrix of the quadratic equation for an ellipse representing the synthesized beam shape on the image plane, s.t., ${\bf x}^{\rm T}{\bf B}{\bf x}=1$. 
The effective source-plane PSFs of the continuum and \oiii{} images shrink by $\approx 40$\% in the east-west direction, whereas they slightly expand by $\approx 3$\% in the north--south direction. 
The respective angular resolutions are found to be $78 \times 84$~mas (PA = $-30\arcdeg$) and $68 \times 76$~mas (PA = $-7\arcdeg$). If assuming $1\farcs 0$ corresponds to 4.7~kpc, the effective spatial resolutions for the dust continuum and \oiii{} images are 370~pc $\times$ 390~pc and 320~pc $\times$ 360~pc, respectively.

Figure~\ref{fig:demagnified} shows the demagnified images of \target{}. The sizes of the dust and UV clumps are typically $\sim 0\farcs 15$, whereas the \oiii{} clumps seem to be slightly smaller ($\sim 0\farcs 1$), implying that the internal structures of this galaxy are on sub-kpc scales. The offsets between the neighboring dust and \oiii{} clumps are $\approx 40$--90~mas or $\approx 200$--400~pc. This is substantially larger than the statistical positional uncertainties of the ALMA images, suggesting the presence of multi-phase ISM within the galaxy.

\subsection{Exclusive spatial distribution of dust and ultraviolet-origin emission}\label{sect:decorrelation}

\begin{figure}[t!]
\includegraphics[width=0.48\textwidth]{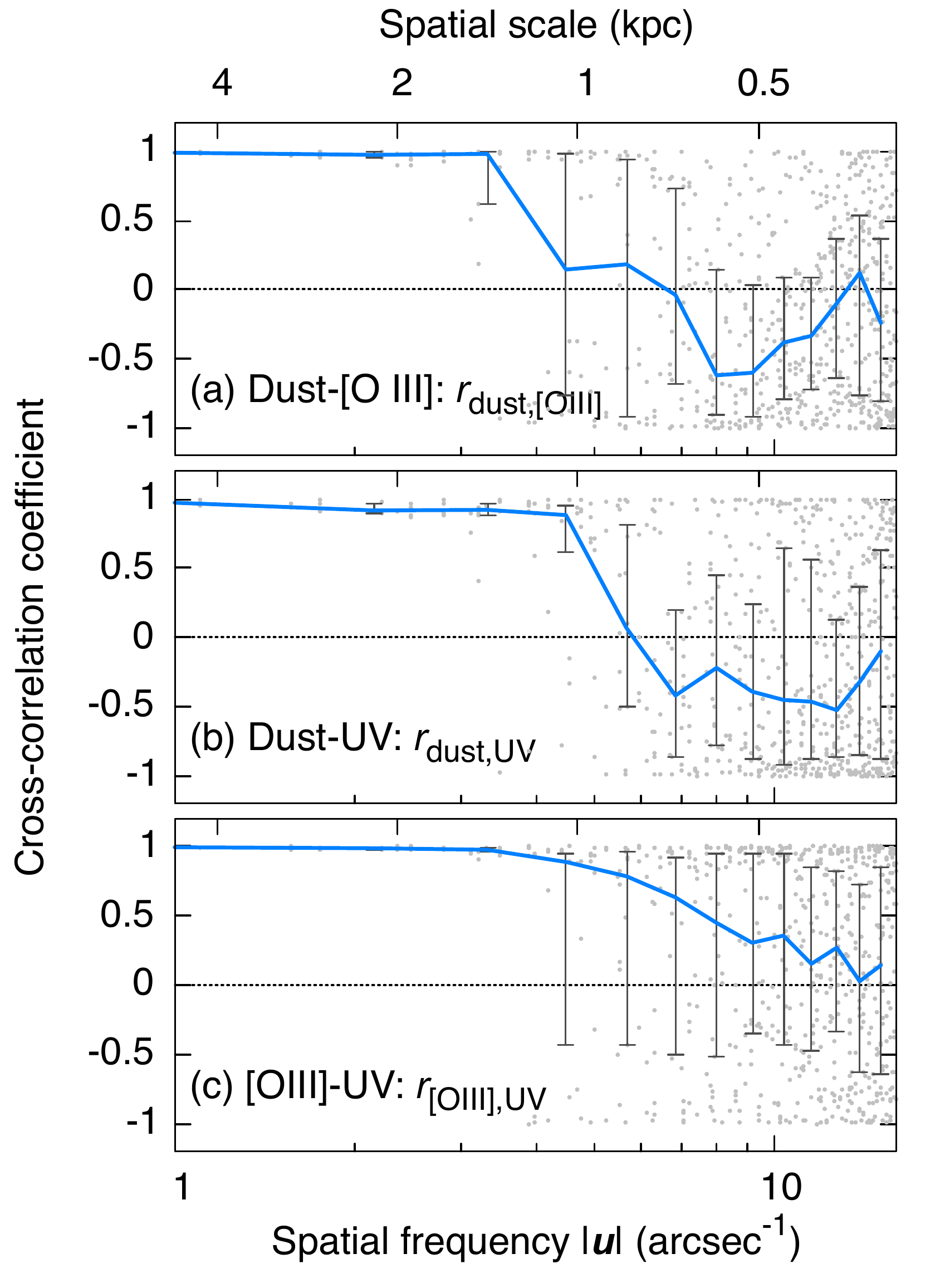}
\caption{
    The cross-correlation coefficient obtained for 3 pairs among images of the 850~$\micron$ dust continuum emission, the 0th moment of the \oiii{} 88~$\micron$ line, and the rest-frame UV continuum emission (WFC3/F160W) obtained for \target{}. The dots show the coefficient, while the blue lines with error bars represent the median coefficient over an annulus at each spatial frequency bin. The error on each plot is taken from the central 50 percentile. The horizontal axis at the top shows the spatial scale, ${\rm 4.7~kpc }\cdot |{\bf u}|^{-1}$. \label{fig:coherence}}
\end{figure}

Generally speaking, all emission of a galaxy is localized in the same volume on a galactic scale, while on smaller scales exclusive spatial distributions are expected between cold dust and UV-origin emission (UV continuum and the \oiii{} line), because rest-frame FIR dust emission predominantly traces the cold neutral/molecular phase whereas UV-origin emission does the warm ionized medium. This means that their spatial distributions in 3-dimensional space are anti-correlated on a certain characteristic scale. If on that spatial scale, relative positions of dust and UV-origin emission can be regarded to be statistically random, their projected (2-dimensional) distributions on the plane of the sky will erase the correlation signal, resulting in no correlation. Here, an angular cross-correlation coefficient is employed to evaluate the spatial scale at which the correlation drops out to zero.

The angular cross-correlation coefficient is expressed as
\begin{eqnarray}
r_{12}({\bf u}) =
\frac{
  \mathcal{F}[I_1({\bf x})]
  \mathcal{F}[I_2({\bf x})]^{\ast}
}{\sqrt{
  |\mathcal{F}[I_1({\bf x})]|^2
  |\mathcal{F}[I_2({\bf x})]^{\ast}|^2
}}\,,
\end{eqnarray}
where $I_i({\bf x})$ is the brightness of the $i$-th image ($i = 1$, 2) at the on-sky position ${\bf x}$, and $\cal F[\cdots]$ is the Fourier transform operator \citep[e.g.,][]{Moriwaki19}. In general $r_{12}$ is a complex function of a 2-dimensional spatial frequency ${\bf u}$. We evaluate the strength of correlation as a function of radius ${\bf |u|}$ on the spatial frequency domain by taking its real part, ${\rm Re}[r_{12}({\bf |u|})]$; e.g., ${\rm Re}[r_{12}] = +1$ and $-1$ indicate positive and negative correlation, and 0 suggests no correlation. Hereafter we refer to ${\rm Re}[r_{12}]$ as the cross-correlation coefficient.

We estimate the cross-correlation coefficients between the three images of the 850~$\micron$ dust continuum emission, the \oiii{} integrated intensity, and the UV continuum emission obtained in the WFC3/F160W band. The images are corrected for gravitational lensing. 
The cross-correlation coefficients are calculated for the central $1\farcs 28 \times 1\farcs 28$ part of the images with a pixel size of 10~mas. The choice of the pixel size does not change the result as far as it is much smaller than the PSF. To reduce shot noise from emission-free pixels, the images are tapered out with a 2-dimensional Gaussian with ${\rm FWHM} = 0\farcs 3$ so that the statistical noise toward the edge of the images is suppressed.
This is equivalent to smoothing with a $|{\bf u}| \sim 3$~arcsec$^{-1}$ width kernel in the spatial frequency domain.

Figure~\ref{fig:coherence} shows the cross-correlation coefficient among the spatial distribution of dust, \oiii{}, and UV emission.  The coefficient is almost unity at a low spatial frequency ($|{\bf u}| \sim 1$~arcsec$^{-1}$). This is likely because the emission is regarded as a point source if we would see it at a $|{\bf u}|^{-1} \sim 1''$ resolution.  The correlation signals are smeared out at high spatial frequencies $|{\bf u}| \gtrsim 15$~arcsec$^{-1}$, likely because the high spatial frequency components are dominated by noise. 
One should also note that $r_{\rm dust, UV}$ and $r_{\rm [OIII], UV}$ at $|{\bf u}| \gtrsim 10$~arcsec$^{-1}$ should be treated with caution because the coarser resolution of the UV image makes the coefficient noisy.
$r_{\rm dust, [OIII]}$ and $r_{\rm dust, UV}$ have clear cutoffs at $|{\bf u}| \approx 4$--6 arcsec$^{-1}$ or $\approx 0\farcs 2$, whereas no apparent cutoff is found in $r_{\rm [OIII], UV}$. This implies that \oiii{} and UV emission are arising from the same regions while dust and \oiii{}/UV are typically separated by $\approx 0\farcs 2$, which corresponds to a physical scale of $\approx 1$~kpc.


\section{Discussions}\label{sec:discussions}

\subsection{Origin of the decorrelation scale}

The cutoff scale we found in the cross-correlation coefficients (the decorrelation scale, Section~\ref{sect:decorrelation}) gives a characteristic scale of different stages of star formation. 
For local and lower-$z$ galaxies, the maximum scale of coherent star formation is about the disk Jeans length. There have been several studies to link gravitational instability to the size scale of massive young stellar clumps found in low-mass star-forming galaxies at lower-$z$ (e.g., \citealt{Elmegreen09a, Elmegreen09b, Dekel09a, Dekel09b}, see also \citealt{Shibuya16} for a review of recent observations). Those studies showed that accretion of cold gas streaming from intergalactic space sustains a high gas fraction and the turbulence necessary for the disk of a galaxy in question to break up into massive clumps by gravitational instability.
If gravitational instability plays a role in determining the decorrelation scale we found, its maximum scale (hereafter, $\ell_{\rm decor}$) should be of the order of the disk Jeans length, although the scale can decrease if the shear motion of a differentially-rotating disk is at work.
In what follows we explore whether the gravitational instability can explain both the decorrelation scale and the high turbulent velocity observed in \target{} by following the discussions of \citet{Elmegreen09b}.

The Jeans length is given by $\ell_{\rm J} \sim \sigma_V^2/\pi G \Sigma$ assuming an infinitely thin disk, where $\sigma_V^2$ is the velocity dispersion, $G$ the gravitational constant, and $\Sigma$ the mass surface density. Let $f_{\rm D} \equiv M_{\rm disk}/M_{\rm halo}$ be a disk-to-halo mass ratio, and then $\pi G \Sigma = V_{\rm c}^2/R(1 + 1/f_{\rm D})$, where $V_{\rm c}$ and $R$ are the circular velocity and radius of the disk, respectively. If the halo contribution to the mass of the galaxy is negligible on a disk scale, i.e., $f_{\rm D} \gtrsim 1$, $\pi G \Sigma \sim V_{\rm c}^2/R$, yielding
$\ell_{\rm J}/R \sim (\sigma_V/V_{\rm c})^2$.

If the decorrelation scale characterizes the typical offset of different stages of star formation in \target{}, gravitational instability emerges on spatial scales $< \ell_{\rm decor}$. Thus, let us assume $\ell_{\rm J} = \ell_{\rm decor}/2 \approx 0.5$~kpc; in fact, this is similar to the clump scale seen in Figure~\ref{fig:demagnified}. Since from Figure~\ref{fig:demagnified} $R = 2$~kpc is a reasonable assumption, $\ell_{\rm J}/R \sim 1/4$, resulting in $\sigma_V \sim V_{\rm c}/2$. This is suggestive of a dispersion-dominated system for \target{}, consistent with the lack of a global velocity gradient in \oiii{}. Although the circular velocity $V_{\rm c}$ is unavailable as \target{} could be face-on (Section~\ref{sec:results}), $V_{\rm c} \sim 50$--100~\kms{} is not an unrealistic value since similar or slightly larger values are found among $z \sim 7$ LBGs with similar or slightly-larger mass scales \citep[e.g.,][]{Smit18, Harikane20, Schouws22}. If this is the case, $\sigma_V \sim 25$--50~\kms{}, consistent with the global velocity dispersion ($\langle \sigma_V \rangle_{\rm w} = 54$~\kms{}, Section~\ref{sec:results}) or those of the individual clumps found in the PVD (Figure~\ref{fig:moment_map}) despite crude estimation.
In other words, the relatively large velocity dispersion of the turbulent gas can naturally be explained if the decorrelation scale is related to gravitational instabilities in the gas. The origins for the large velocity dispersion of pre-star-forming gas are currently unknown, but possible origins include accretion of pristine gas from outside the galaxy and/or kinetic energy injection from the past stellar feedback.

Another interpretation might be a merger of smaller dwarf galaxies.
One of the well-studied dwarf mergers in the local Universe that is comparable to \target{} is a local blue compact galaxy Haro~11. Haro~11 is similar in size ($\sim 4$~kpc), SFR ($\sim 20$--30 $M_\Sol$~yr$^{-1}$), and metallicity ($\sim 0.3 Z_\Sol$) to \target{}, although the stellar and dust masses of Haro~11 are an order of magnitude larger  (\citealt{Ostlin15}, and references therein. See also \S~\ref{sec:introduction} for the physical properties of \target{}). Haro~11 has three knots; two cores of stars and a group of bright stellar clusters. The knots are separated by $\sim 1$~kpc and are bridged by dust lanes, which is apparently similar to \target{}. Haro~11 has two diffuse stellar disks overlapped with each other. It also shows disturbed features which are often seen in interacting galaxies, such as tidal arms and discontinuity in brightness of the stellar disks. Integral field unit and slit spectroscopy of ionized gas in the optical also revealed multiple kinematical components of Haro~11, each of which is associated with the knots, strongly suggesting the evidence for a merger of two progenitor systems \citep{Ostlin15}.
A similar situation is also observed in bright LBGs at high redshifts, such as a $z = 7.15$ LBG, B14-65666 (a.k.a. the Big Three Dragons). B14-65666 is detected in \oiii{}, \cii{}, and dust and is similar in stellar properties (e.g., SFR, stellar mass and metallicity) to \target{}. It shows two distinct clumps with angular and velocity offsets of $\sim 3$~kpc and $177 \pm 16$~\kms{}, respectively, strongly indicating an ongoing merger \citep{Hashimoto19}.

\target{} shows, however, no distinctive velocity components implying merger remnants at the positions of the stellar clumps (see Figure~\ref{fig:moment_map}). The velocity offsets among the stellar or \oiii{} clumps are negligible, so we conclude that the clumpy morphology is not likely due to a merger but represents huge star-forming regions embedded in a single galaxy. Future NIR imaging with \jwst{} will be able to reveal the morphology.

\subsection{Dust cavity: A possible superbubble?}\label{sec:superbubble}

As we have seen in Section~\ref{sec:results}, \target{} has a long asymmetric tail in which a cavity is evident in the dust continuum image. Although the significance is not very high, this can be possible evidence for ongoing SN feedback.
Interestingly, recent hydrodynamic simulations of galaxy formation predict a bubble-like structure likely created by a starburst, which is detectable in a 850~$\micron$ continuum image \citep{Arata19}.
In this section, we discuss a possibility that the dust cavity is created by SN feedback driven by ongoing star formation, in terms of energy budget for creation of a superbubble.

As SNe and stellar winds are thought to play a substantial role in regulating subsequent star formation, many theoretical models of galaxy formation have taken them into account \citep[e.g.,][]{Arimoto87}. The latest hydrodynamic simulations of galaxies in the EoR reveal that in the earliest era of galaxy formation, SN feedback instantaneously blows off the star-forming clouds because the shallow potential allows the gas to escape, resulting in episodic change in SFR \citep{Katz19,Arata20}. Indeed, recent SED analyses of $z = 7$--9 galaxies including \target{} revealed a flux excess in the rest-frame optical bands.  The attribution could be the past star formation activity that had suddenly been terminated by SN feedback at the peak of star formation \citep{Hashimoto18, Hashimoto19, Tamura19}.

One of the most pronounced forms of SN feedback in low-mass galaxies is a superbubble followed by galactic winds. In starburst galaxies, a series of SNe supplies almost constant kinetic energy into the surrounding ISM, and cavities created by individual SNe merge with each other, eventually making a galactic scale cavity called a superbubble. It expands toward low density regions, typically in the vertical direction with respect to a galactic disk, and finally leads to a galactic-scale outflow \citep[e.g.,][]{Tomisaka86}.

Whether energy injection from a series of SNe makes an interstellar bubble and eventually leads to a galactic-scale outflow is dependent on the relative importance of the kinetic energy pushing the surrounding media to create the bubble and radiative cooling of the gas within the bubble. According to the formulation by \citet{MacLow88} \citep[see also][]{Recchi01}, the radiative cooling time-scale is expressed as $t_{\rm R} \sim 16(\beta Z)^{-35/22}L_{38}^{3/11}n^{-8/11}$~Myr, where $\beta$ is a constant of order of unity, $Z$ is the metallicity in solar units, $L_{38}$ is the mechanical luminosity of stellar feedback in units of $1 \times 10^{38}$~erg~s$^{-1}$, and $n$ is the number density of surrounding medium. The cavity radius when time reaches $t_R$ is $R_{\rm R} \sim 350 (\beta Z)^{-27/22}L_{38}^{4/11}n^{-7/11}$~pc. On the other hand, the dynamical time-scale is characterized as a time-scale for which the bubble radius reaches the scale height of a galaxy and is expressed as $t_{\rm D} \sim H^{5/3} (\rho/L_{\rm w})^{1/3}$, where $H$ and $\rho$ are the scale height and the gas density of the surrounding medium. $L_{\rm w} \equiv 0.5 \dot{M}V_{\rm w}^2$ is the mechanical luminosity of the wind, where $\dot{M}$ and $V_{\rm w}$ are the mass loss rate and wind velocity, respectively. If the ratio of radiative cooling to dynamical time-scales,
\[
\frac{t_{\rm R}}{t_{\rm D}} \sim 
8.22\, n^{-35/33} L_{38}^{20/33}
\left(\frac{H}{\rm 100\ pc}\right)^{-5/3}
(\beta Z)^{-35/22}\,,
\]
is greater than unity, the cavity reaches the scale-height of the galactic disk, resulting in the occurrence of galactic outflow in the dynamical timescale.

For the cavity of \target{} with reasonable assumptions, the cooling time-scale is likely to be longer than the dynamical time-scale of a cavity. This means that a 1-kpc scale cavity can occur.  We assume that the kinetic energy of a single SNII is approximately 10\% of the total energy, i.e., $E_{\rm kin} = 0.1 \times E_{\rm SN} = 1 \times 10^{50}$~erg. Since the supernova rate per unit SFR is quite uncertain at the EoR, we just assume the value derived from the \citet{Salpeter55} initial mass function ($0.010~M_\Sol^{-1}$, e.g., \citealt{Inoue11}). Note that the choice of initial mass function does not largely change the supernova rate; it becomes $0.017~M_\Sol^{-1}$ if we choose the \citet{Chabrier03} mass function, for instance. \citet{Tamura19} showed that the global SFR and metallicity of \target{} are SFR $\sim 100~M_\Sol$~yr$^{-1}$ and $Z \approx 0.2~Z_\Sol$, respectively.  We assume that only a small fraction (1\%) of the global SFR contributed to the formation of the cavity, while the metallicity is the same as the global one. We set the scale height to be the radius of the cavity (75~mas or 350~pc). The surrounding medium is likely to be a mixture of warm neutral medium (WNM), cold neutral medium (CNM), and ionized medium with a volume-weighted ISM density of $n = 10$~cm$^{-3}$ for a fiducial value. Note that for the propagation of the SN wind a volume-weighted density is important; WNM is considered to dominate the volume, the mean gas density would be close to that of WNM ($\sim 0.1$--1~cm$^{-3}$ for the local Universe), so $n = 10$~cm$^{-3}$ is considered to be an upper boundary. This yields $t_{\rm R}/t_{\rm D} \sim 40 \gg 1$ and $t_{\rm D} \sim 5$~Myr, suggesting that the SN feedback from a star-forming region with a tiny fraction of the global SFR can sweep up the surrounding diffuse media to form a 1~kpc superbubble in a time-scale comparable to the age of the star forming region.

Here we want to point out a few caveats. Firstly, in the \oiii{} PVD (Figure~\ref{fig:moment_map}f) no clear expanding shell is found at the position of the dust cavity. If an expanding bubble is located in a static, uniform medium which emits a spectral line, it would appear as an annulus in the PVD of the emission line, but Figure~\ref{fig:moment_map}f does not show such a structure.  However, an expanding structure does not necessarily show an annulus in the PVD seen in forbidden lines. Imaging spectroscopy of local giant \hii{} regions, where 10 to 100~pc scale superbubbles are found, reveals no clear annulus in their PVD but a velocity discontinuity or jump across the superbubbles \citep{Melnick99, Torres-Flores13}.  In addition, potential contaminants of ionized gas which are not associated with the bubble but are located along the line of sight can make the characteristic velocity structure of the bubble unclear.
Secondly, the stellar age estimated from the SED fits \citep[$\approx 4$~Myr,][]{Tamura19} is too young to produce core-collapse SNe with the progenitor masses $\lesssim 60~M_\Sol$. 
This 4~Myr period is so stringent that the parental dense clouds of gas surrounding the progenitors of SNe would hamper the initial expansion of the SN shocks, which could take $> 4$~Myr for parental cloud dispersal \citep{Sommovigo20}.
This apparent conflict could be mitigated without any substantial conflict with the stellar SED and thus the inferred age, if the stellar component has a small ($\sim 1\%$) subset which has been keeping star formation to create the superbubble for a time-scale comparable to the lifetime of a $8 M_\Sol$ star (50~Myr).
It is also possible that the SNe formed either in a low density environment, close to the edge of the parental cloud, or in a region where radiation and winds of their progenitors have processed and dissipated the dense gas \citep{Decataldo20}, allowing the bubble to expand rapidly.

Our calculations indicate that the dust cavity can be a galactic scale superbubble driven by only a small fraction of the global star formation. It implies that there is much more sub-galactic scale feedback than that suggested from the PVD (Figure~\ref{fig:moment_map}). 
SN feedback would not only affect the interstellar medium but also produce outflows, eventually contributing the metal enrichment in circumngalactic or intergalactic space. This could be the origins of extended \cii{} emission recently found around high-redshift LBGs \citep[e.g.,][]{Fujimoto20}.
SN feedback can also change the ionization structure of the ISM and thus plays an important role in regulating ionizing photon escape. SN explosions create many cavities in the neutral media and fill them with ionized gas, which makes the neutral ISM more porous and reduces the covering fraction of neutral ISM surrounding \hii{} regions. Indeed, recent observations of nebular emission lines from the rest-frame UV to FIR have revealed reduced covering fractions in local dwarf galaxies \citep{Cormier19} and in $z \sim 6$ LBGs \citep{Harikane20, Sunaga20}.  The possible porous geometry will help ionizing photons from massive stars escape outside the galaxy and could eventually contribute the cosmic reionization, although the SED modeling for the entire galaxy prefers zero escape of ionizing photons (the escape fraction of $f_{\rm esc} = 0.00^{+0.19}_{-0.00}$ for the \citet{Calzetti00} extinction law, \citealt{Tamura19}).


\section{Conclusions}\label{sec:conclusions}

We report the results of $< 100$ mas ALMA imaging of the \oiii{} 88~$\micron$ line and dust continuum emission from a $z = 8.312$ Lyman break galaxy \target{}.
The integration times in  \oiii{} and 850~$\micron$ continuum are 18.8 and 27.5 hr, respectively. The intrinsic spatial resolution corrected for gravitational lensing is $\approx 300$~pc, providing the first and furthest imaging resolving multi-phase ISM in a galaxy in the middle of the epoch of reionization.

The velocity-integrated \oiii{} emission has three peaks which are likely associated with three young stellar clumps of \target{}, while the channel map shows a very complicated velocity structure with little global velocity gradient unlike what was found in the lower-resolution ($\sim 0\farcs 5$) \cii{} 158~$\micron$ image, suggesting random bulk motion of ionized gas clouds in the galaxy. In contrast, dust emission has two individual clumps apparently separating or bridging the \oiii{}/stellar clumps. The cross correlation coefficient between the dust and \oiii{}/ultraviolet emission is unity on a galactic scale, while it drops at $< 1$~kpc, suggesting well mixed geometry of multi-phase interstellar media on this spatial scale. If the cutoff scale characterizes different stages of star formation, the cutoff scale can be explained by gravitational instability of turbulent gas. The dust emission shows a fairly asymmetric distribution in the direction perpendicular to the stellar extent with galaxy-wide outskirts extending outside the stellar disk. Furthermore, an off-center cavity is embedded in the dust continuum image. This could be a superbubble producing galactic-scale outflows, since energy injection from only a small fraction of the star formation lasting several~Myr is large enough to push the surrounding media to make a kpc-scale cavity.

These facts suggest that \target{} could be in an early phase of stellar feedback. A new insight into the exact location of the stellar components and detailed gas kinematics traced out by ionized and neutral media will be of great help to understand the stage of stellar feedback. Upcoming high-resolution \cii{} imaging with ALMA (Bakx et al., in preparation) and near-infrared imaging and spectroscopy with \jwst{} will allow us to investigate this further.

\begin{acknowledgments}
We acknowledge the referee for detailed and elaborate scientific comments. YT thanks Kouichiro Nakanishi, Kotaro Kohno, and Yoshinobu Fudamoto for fruitful discussions. 

This study is supported by NAOJ ALMA Scientific Research Grant Number 2018-09B and JSPS KAKENHI (No.\ 17H06130, 19H01931, 20H01951, 20H05861, 20K22358, 21H01128, 21H04496, 22J21948, 22H01258, and 22H04939). 
AKI and YS are supported by NAOJ ALMA Scientific Research Grant Number 2020-16B.
TH was supported by Leading Initiative for Excellent Young Researchers, MEXT, Japan (HJH02007). 
EZ acknowledges funding from the Swedish National Space Agency.
TTT was supported in part by the Collaboration Funding of the Institute of Statistical Mathematics ``New Perspective of the Cosmology Pioneered by the Fusion of Data Science and Physics''. 

This paper makes use of the following ALMA data: 
ADS/JAO.ALMA \#2016.1.00117.S, 
ADS/JAO.ALMA \#2017.1.00225.S, 
ADS/JAO.ALMA \#2017.1.00486.S, and 
ADS/JAO.ALMA \#2018.1.01241.S.
ALMA is a partnership of ESO (representing its member states), NSF (USA) and NINS (Japan), together with NRC (Canada), MOST and ASIAA (Taiwan), and KASI (Republic of Korea), in cooperation with the Republic of Chile. The Joint ALMA Observatory is operated by ESO, AUI/NRAO and NAOJ.
This work has made use of data from the European Space Agency (ESA) mission {\it Gaia} (\url{https://www.cosmos.esa.int/gaia}), processed by the {\it Gaia} Data Processing and Analysis Consortium (DPAC, \url{https://www.cosmos.esa.int/web/gaia/dpac/consortium}). Funding for the DPAC has been provided by national institutions, in particular the institutions participating in the {\it Gaia} Multilateral Agreement.
Data analysis was in part carried out on the Multi-wavelength Data Analysis System operated by the Astronomy Data Center (ADC), National Astronomical Observatory of Japan.

\end{acknowledgments}

%

\facilities{ALMA, Gaia, HST(WFC3)}


\software{astropy \citep{2013A&A...558A..33A},  
          CASA \citep{McMullin07}, 
          IRAF \citep{Tody93},
          Miriad \citep{Sault95}
          }






\bibliography{sample63}{}
\bibliographystyle{aasjournal}



\end{document}